\documentclass[%
	 preprint,
	  amsmath,amssymb,
	   aps,longbibliography
	   ]{revtex4-2}

\usepackage{multirow}
\usepackage{numprint}
\usepackage{siunitx}
\usepackage{graphicx}
\usepackage{dcolumn}
\usepackage{bm}

\usepackage[utf8]{inputenc}
\usepackage[T1]{fontenc}
\usepackage{mathptmx}
\usepackage{etoolbox}
\usepackage{adjustbox}
\makeatletter
\def\@email#1#2{%
 \endgroup
 \patchcmd{\titleblock@produce}
  {\frontmatter@RRAPformat}
  {\frontmatter@RRAPformat{\produce@RRAP{*#1\href{mailto:#2}{#2}}}\frontmatter@RRAPformat}
  {}{}
}%
\makeatother
\begin{document}

\title[Sample title]{Acceptor and donor impurity levels in hexagonal-diamond silicon}
\author{Marc T\'unica}
\affiliation{Université Paris-Saclay, CNRS, Laboratoire de Physique des Solides, 91405 Orsay, France}
\author{Alberto Zobelli}
\affiliation{Université Paris-Saclay, CNRS, Laboratoire de Physique des Solides, 91405 Orsay, France}
\author{Michele Amato$^{*}$}
\affiliation{Université Paris-Saclay, CNRS, Laboratoire de Physique des Solides, 91405 Orsay, France}
\email{michele.amato@universite-paris-saclay.fr}

\date{\today}

\begin{abstract}
Recent advances in the characterization of hexagonal-diamond silicon (2H-Si) have shown that this material possesses remarkably different structural, electronic and optical properties as compared to the common cubic-diamond (3C) polytype. Interestingly, despite the wide range of physical properties analyzed, to date no study has investigated impurity energy levels in 2H-Si. Here, we present results of ab initio DFT simulations to describe the effect of p- and n-type substitutional doping on the structural and electronic properties of hexagonal-diamond Si (2H-Si). We first provide a detailed analysis of how a given impurity can assume a different local symmetry depending on the host crystal phase. Then, by studying neutral and charged dopants, we carefully estimate donors and acceptors transition energy levels in 2H-Si and compare them with the cubic-diamond (3C) case. In the case of acceptors, the formation energy is always lower in 2H-Si and is associated with a shallower charge transition level with respect to 3C-Si. On the other hand, donors prefer the cubic phase and have transition energies smaller with respect to 2H-Si. Finally, by employing a simple model based on the 2H/3C band offset diagram, we prove the physical validity of our findings and we show how holes can be used to stabilize the 2H-Si phase. Overall, the described doping properties represent a robust starting point for further theoretical and experimental investigations.
\end{abstract}

\maketitle

\section{\label{sec:introduction}Introduction}
Recent developments and advances in semiconductor growth methods demonstrated that the hexagonal-diamond polytype of Si, Ge and SiGe alloys (2H in the Rasmdell notation~\cite{Ramsdell1947}) can be fabricated in the form of nanowires~\cite{GalvaoEPJB2020,FadalyNATURE2020,VincentNL2015,TangNANOSCALE2017,HaugeNL2015}. This crystal structure presents a $C_{3v}$ symmetry (P6$_{3}$/mmc space group) with three equal length bonds in the c-plane (0001) and a larger one along the [0001] direction of the hexagonal crystal cell. Compared to the 3C cubic-diamond phase (Fd-3m group symmetry) where four equal length bonds induce a $T_{d}$ symmetry, in the 2H phase the atomic ordering along the $\langle 111 \rangle$ cubic direction changes from an ABC to an ABAB stacking. 

The main reason for studying these novel group IV polytypes lies in the possibility to observe unexplored electronic and optical properties with respect to those of their 3C counterparts. To give an example, Ge, which in its cubic form is an indirect semiconductor, has a direct band gap in its hexagonal polytype~\cite{RodlPRM2019,KaewmarayaJPC2017}. Yet, hexagonal SiGe alloys~\cite{HaugeNL2017} and quantum wells~\cite{PeetersNATCOMM2024} have recently attracted a great attention because of their direct band gap and the experimentally observed light emission that can be tailored via the Ge concentration and quantum confinement effects~\cite{FadalyNATURE2020,BorlidoPRM2021,PeetersNATCOMM2024}. In the case of Si, while the cubic phase has an indirect band gap of 1.12~eV with a conduction band minimum close to the X point, the 2H structure presents an indirect band gap of about 0.95~eV and a conduction band minimum at the L point~\cite{RodlPRB2015,KellerPRM2023}. The direct band gap is also reduced from 3.40~eV to 1.63~eV~\cite{RodlPRB2015}. Indeed, 2H-Si presents a larger optical absorption spectra in the visible range than 3C-Si~\cite{AmatoNL2016}, making it a good candidate for solar panel materials. Furthermore, 3C/2H Si junctions are also of interest for photovoltaics due to their intrinsic tendency to promote charge carrier separation~\cite{AmatoNL2016,KaewmarayaJPC2017,BelabbesPRB2022}. 

It is well known that most technologies based on carrier transport in semiconductors (e.g. electronics, embedded systems, light emitting devices and solar cells) become functional only if the material can be appropriately doped. Impurities play a crucial role as they allow to modulate the carrier concentration around the Fermi level. Furthermore, in different type of semiconductor compounds, doping can be employed to stabilize one polytype with respect to the others~\cite{HeineJACS1991,DalpianAPL2006,DalpianPRL2004,LiuAPL2024}. For instance, Cai et al.~\cite{CaiPRB2021} recently demonstrated that the direct band gap 2H-Ge phase can be stabilized by incorporating electrons into the 3C-Ge crystal. Finally, most of the experimental grown nanowires can contain significant concentration of extrinsic dopants due to the substrate atomic contamination. This is, for instance, the case of 2H-Ge and 2H-SiGe alloys~\cite{FadalyNATURE2020} fabricated with the crystal transfer methods which usually present a high concentration of As atoms (~$\sim$~10$^{18}$~cm$^{-3}$).  

Although the behaviour of impurities in 3C-Si has been widely discussed, to date no study has investigated impurity energy levels in 2H-Si. The experimental characterization of dopants in this polytype still remains challenging because of the difficulties in its fabrications methods. Ab-initio theoretical approaches can be hence an efficient way to investigate this question. A few studies focused on the analysis of the stability of single-point and neutral substitutional defects in 2H-Si~\cite{AmatoNL2019,AmatoJPC2020,SunPRM2021} but they were essentially limited to the structural stability of defects. 
Here, we present a comprehensive study carried out via ab initio DFT simulations on the effect of p- and n-type substitutional defects on the structural and electronic properties of 2H-Si. We first provide a detailed analysis on how a given impurity can assume a different local symmetry depending on the host crystal phase. Then, by studying neutral and charged dopants, we carefully estimate donors and acceptors thermodynamic transition levels in 2H-Si and compare them with the 3C-Si case. Finally, by employing a simple model based on the 2H/3C band offset diagram, we show how the carrier concentration can be used to stabilize one phase with respect to the other.

\section{\label{sec:methodology}Methodology}
All simulations were performed by spin-polarized ab initio DFT as implemented in the SIESTA code~\cite{SolerJCMP2002}, using the generalized gradient approximation (GGA) with the Perdew, Burke, Ernzerhof (PBE) functional~\cite{PerdewPRL1996}. Since the diameter of the 2H nanowires grown experimentally is usually larger than 100 nm, we considered bulk supercells as an accurate and reliable approximation for such systems. Indeed at such large diameters, quantum confinement effect on the band gap can be neglected~\cite{AmatoMSEB2012,AmatoPSSB2010,MohammadJPCM2014}. Norm-conserving pseudopotentials of Trouillier-Martins type were chosen to replace the core electrons. Valence electrons were instead represented using a double-$\zeta$ (DZ) polarized basis set for Si, and double-$\zeta$ plus two polarization orbitals (DZP) basis sets for impurities. We considered p-type (B, Al, Ga, In) and n-type (P, Sb, As) as common substitutional dopants in 2H-Si. Lattice parameters and band gap of pristine bulk 3C- and 2H-Si were calculated by sampling the Brillouin zone using a 8$\times$8$\times$8 Monkhorst-Pack k-points grid~\cite{MonkhorstPRB1976}. Table~\ref{tab:table_pure_si} presents the converged results of the lattice parameter and the band gap calculations for both the phases as well as their experimental values. As expected, when compared to experiments, it is clear that both quantities are affected by the error associated to the GGA exchange-correlation functional. We used a denser grid of 25$\times$25$\times$25 k-points to calculate the dielectric constant of both phases. This quantity was needed for the charge correction scheme described below~\cite{FreysoldtPRL2009}. The values of the in-plane and out-of-plane components, reported in Table~\ref{tab:table_pure_si}, are in perfect agreement with other DFT calculations~\cite{BaroniPRB1986, ZhengPRB2017} and only slightly overestimate the experimental values~\cite{Li1980}. For the doped systems, a supercell of 512 (a 4$\times$4$\times$4 replica of the 8-atoms conventional supercell with orthogonal cell vectors) and of 500 atoms (a 5$\times$5$\times$5 replica of the primitive cell) were used for the 3C and 2H phases, respectively. This choice has guaranteed the convergence of the formation energy value for both neutral and charged impurities as it eliminated the spurious interaction between defect periodical images (see Supplemental Materials (SM)~\cite{supp}). Such doped supercells correspond to a dopant concentration between~10$^{19}$~cm$^{-3}$ and~10$^{20}$~cm$^{-3}$, a value that is one order of magnitude larger with respect to the impurity concentration usually measured in 2H-samples~\cite{FadalyNATURE2020}. A 2$\times$2$\times$2 k-grid and a real space grid cut-off of 500~Ry were considered to ensure the proper convergence of the results. Self-consistency field cycles were converged with a tolerance of $10^{-5}$~eV. Atomic positions were optimized until the forces were less than $10^{-2}$~eV/\AA~and the stress lower than $10^{-1}$~GPa. Relaxation was performed adopting a conjugate gradient algorithm that calculates DFT forces through the Hellman-Feynman theorem. Charged structures were relaxed starting from the neutral defect optimized structure. A compensating jellium background was added to avoid divergences due to the defect-defect long-range electrostatic interaction between charged periodic cells. 

For all the considered impurities, we calculated their formation energy, $E_{form}$, in a given crystal phase (2H or 3C). This quantity is defined as the energy required to replace a host atom with a neutral or charged defect. In the Zhang and Northrup formalism~\cite{ZhangPRL1991}, it reads as follows:
\begin{equation}\label{eq:formation_energy}
E_{form} = E_{total}^{D} - E_{total}^{prist} + \mu_{0} - \mu_{D} + q (\epsilon_{VBM}+\mu_{e}) + E_{corr},
\end{equation} 
where $E_{total}^{D}$ is the total energy of the doped system, $E_{total}^{prist}$ is the total energy of the pristine system, $\mu_{0}$ is the chemical potential of Si in a given phase while $\mu_{D}$ is that of the defect. $\mu_{0}$ is here defined as the total ground energy per atom of a pristine Si cell whereas $\mu_{D}$ is calculated as the total energy of the isolated impurity atom. However, it is worth to note that there is no univocal criteria to calculate the chemical potential of an impurity as this quantity strongly depends on the synthesis conditions. As a consequence, the calculated value of the formation energy can assume a wide range of values depending on the way $\mu_{D}$ is defined. Nevertheless, since we are interested in the relative stability of a dopant among the two phases, the chemical potential $\mu_{D}$ erases if we refer to the difference between formation energies, $\Delta E^{2H-3C}_{form} = E_{form}^{2H} - E_{form}^{3C}$ rather than in $E_{form}$ value in itself (this also holds because the difference of $\mu_{0}$ among the two phases can be neglected). Yet, referring back to Eq.~\eqref{eq:formation_energy}, the term $q (\epsilon_{VBM}+\mu_{e})$ is the charge multiplied by the electron chemical potential, $\mu_{e}$, using as energy reference the valence band maximum, $\epsilon_{VBM}$. $\mu_{e}$ varies from 0 to the electronic band gap. Finally, the term $E_{corr}$ corresponds to the Freysoldt-Neugebauer-Van de Walle's finite-size correction~\cite{FreysoldtPRL2009} that cancels artificial interactions between charged defects and that has been proven to give very accurate results for different type of defects~\cite{FreysoldtRMP2014}. As other proposed schemes~\cite{DaboPRB2008,KumagaiPRB2014,FreysoldtRMP2014,NaikCPC2018}, this approach implies the solution of the Poisson equation for a model system and the alignment of the electrostatic potentials to derive the formation energy correction. In particular, solving the Poisson equation for the periodic model potential needs the calculation of the dielectric tensor profile of the material~\cite{NaikCPC2018}, whose static components are presented in Table~\ref{tab:table_pure_si}. 

By calculating formation energies, we can also estimate the thermodynamic transition level between two charge states, $\epsilon$($q_1$/$q_2$), which corresponds to the specific value of the electron chemical potential for which the two states have the same value of $E_{form}$. This quantity can be calculated as:
\begin{equation}\label{eq:transition_level}
\begin{aligned}
\begin{split}
   & \epsilon(q_1/q_2) = \dfrac{E^{D}_{total}(q_1) + E_{corr}(q_1)- E^{D}_{total}(q_2) - E_{corr}(q_2)}{q_2-q_1} \\
   & - \epsilon_{VBM}.
\end{split}
\end{aligned}
\end{equation} 

For each dopant the formation energy plots are reported for different values of $q$ (-1, 0 and +1) and they are linear functions of the chemical potential as shown in Eq.~\ref{eq:formation_energy}. The intersection of the formation energy lines for two different charge states $q_1$ and $q_2$ is the thermodynamic transition level calculated through Eq.~\ref{eq:transition_level}

Transition levels $\epsilon(q_1/q_2)$ correspond to thermal ionization energies that can be measured by deep level transient spectroscopy (DLTS)~\cite{Willardson1998}. We employed Eq.~\eqref{eq:transition_level} to calculate both acceptors, $\epsilon_{A}(q_1/q_2)$ and donor transition levels $\epsilon_{D}(q_1/q_2)$. It is worth to note that transition levels are not affected by the dopant chemical potential which assumes the same value for the charged and neutral impurity and hence cancels out in Eq.~\eqref{eq:transition_level}. 

In the following analysis, Eq.~\eqref{eq:transition_level}, which depends on the valence band maximum of each phase taken separately, allows to compare the doping with a given impurity of the 2H and the 3C phase of Si. On the other hand, to give an accurate quantitative estimation of transition levels, in particular in the case of the phase coexistence~\cite{VincentNL2015,TangNANOSCALE2017,AmatoNL2016}, we align $\epsilon$($q_1$/$q_2$) to a common reference level, i.e. the vacuum level $\Phi$ (see SM for the details of the calculations~\cite{supp}). Indeed, many works have demonstrated that, even if within a semi-local approach, energy levels of dopants generally remain preserved if they are referred to a common well defined potential~\cite{AlkauskasPRB2011,KomsaPRB2010,AlkauskasPRL2008}. This choice improves the quantitative comparison among the two phases. On the basis of this consideration, Eq.~\eqref{eq:transition_level} can be modified as:

\begin{equation}\label{eq:transition_level_phi}
\begin{aligned}
\begin{split}
& \epsilon(q_1/q_2) = \dfrac{E^{D}_{total}(q_1) + E_{corr}(q_1)- E^{D}_{total}(q_2) - E_{corr}(q_2)}{q_2-q_1} \\
& - (\epsilon_{VBM} - \Phi) = \overline{\epsilon}(q_1/q_2) - \overline{\epsilon}_{VBM}.\\
\end{split}
\end{aligned}
\end{equation} 

where the charge transition level $\overline{\epsilon}(q_1/q_2)$ and the $\overline{\epsilon}_{VBM}$ are referred to the vacuum potential. 

\section{\label{sec:structural analysis}Structural properties}
We start our discussion by a detailed structural analysis of the relaxed doped structures. Table~\ref{tab:acceptors_geometry} and \ref{tab:donors_geometry} report the first-neighbors bond lengths of group III and group V impurities in the cubic and hexagonal crystals. In the 3C-Si, both neutral donors and acceptors maintain the tetrahedral symmetry ($T_{d}$) forming four first-neighbor bonds of equal length. Variations in bond lengths compared with the pristine system are uniform and depend on the mismatch between the covalent radius of the impurity and that of the host element, as already reported by Amato et al.~\cite{AmatoNL2019,AmatoJPC2020}. The bond contractions and expansions discussed here agree with previous results. As for smaller defects, such as B and P, the first neighbors impurity bonds are compressed in respect to those of the pristine cell (see first block of Table~\ref{tab:acceptors_geometry} and ~\ref{tab:donors_geometry}) while for all the other reported defects (Al, Ga, In, Sb and As in Table~\ref{tab:acceptors_geometry} and ~\ref{tab:donors_geometry}) they tend to expand since their covalent radius is larger than those of Si. More in general, we can say that impurities in the 3C-Si phase induce an expansion (contraction) of their first neighbors bonds and, consequently, there is a noticeable contraction (expansion) between the first and second neighbors atomic layers that gradually fades within the crystal. The charged impurities also preserve the cubic $T_{d}$ symmetry. As for B$^{-}$, the extra electron does not induce any appreciable change in the bond distances (see first block of Table~\ref{tab:acceptors_geometry}). Al$^{-}$, Ga$^{-}$, and As$^{-}$ first neighbors distances are only slightly reduced compared to the neutral case (see Table~\ref{tab:acceptors_geometry}). These tiny bond length variations are proportional to the impurity size, being more pronounced for larger dopants. Therefore, adding an extra electron to large radius acceptors (Al, Ga, In) in the 3C-Si host lattice induces a larger contraction compared to the neutral case, presumably because the lattice has fewer degrees of freedom to adjust to the extra electron configuration, resulting in a more significant contraction. The analysis of donors shows that, when positive charges are added, bonds tend to contract as well, but without any clear general trend (see Table~\ref{tab:donors_geometry}).\\

Compared to the 3C-Si case, the 2H-Si host lattice presents a lower symmetry due to the anisotropy along the [0001] hexagonal direction ($C_{3v}$ symmetry). In the pristine structure, for a given Si atom, three equal length first-neighbor bonds in the c-(0001) plane can be defined. The fourth bond is longer and perpendicular to the plane, breaking the tetrahedral symmetry. This arrangement of the atoms increases the degrees of freedom for the structural relaxation around the defect. As shown in the first block of Table~\ref{tab:acceptors_geometry}, the local symmetry of the B$^{0}$ impurity is $C_{3v}$: there exist three equivalent bonds in the plane and a longer one along the [0001] crystal direction. This is a consequence of three valence electrons of the acceptor trying to keep the three-fold coordination. The difference between the average of the three in-plane bonds and the out-of-plane one is larger than that of the 2H pristine case. As for B$^{-}$, we observe the same symmetry but slightly more pronounced: an equivalence of the three bonds in the plane and a slight expansion of the perpendicular one with respect to the neutral case. Looking at the second block of Table~\ref{tab:acceptors_geometry}, we can recognize a similar behavior in the case of Al:  both Al$^{0}$ and Al$^{-}$ keep a clear $C_{3v}$ symmetry with three equal length bonds in the plane and a longer one along the $c$ hexagonal direction. As for Ga$^{0}$ and In$^{0}$ (third and fourth blocks of Table~\ref{tab:acceptors_geometry}), though it is not very clear, it seems that the local symmetry is closer to the $T_{d}$ cubic one. Two bonds in the plane are larger than the third one, and even larger than the perpendicular interatomic distance, with this effect being more pronounced for In due to its larger atomic size. When a charge is added, the tendency toward $T_{d}$ symmetry becomes more evident, especially for In (fourth block of Table~\ref{tab:acceptors_geometry}). Possible factors that could explain this deviation from the expected bond length pattern include local distortions, impurity size mismatch as well as electronic wave function symmetry breaking. 
On the other hand, donors in 2H-Si induce similar contraction and expansion of the local structure, but with a different impurity symmetry. As reported in the first block of the Table~\ref{tab:donors_geometry}, P$^{0}$ has a $C_{3v}$ symmetry with a difference between the in-plane and the out-of-the-plane bond lengths larger than those of the pure 2H-Si. When a charge is added, the out-of-plane bond length is slightly reduced, while all the in-plane bonds maintain the same value. There is almost no change in the symmetry between neutral and charged cases, keeping a pronounced $C_{3v}$ symmetry. The same behavior can be recognized also for As and Sb (second and third block of Table~\ref{tab:donors_geometry}). It is worth to note that the difference between the in-plane and the out-of-plane bond length average is strictly related to the impurity ionic radius being lower for Sb and larger for P.\\

\section{\label{formation}Formation Energy}
\begin{figure}
\includegraphics[width=\columnwidth]{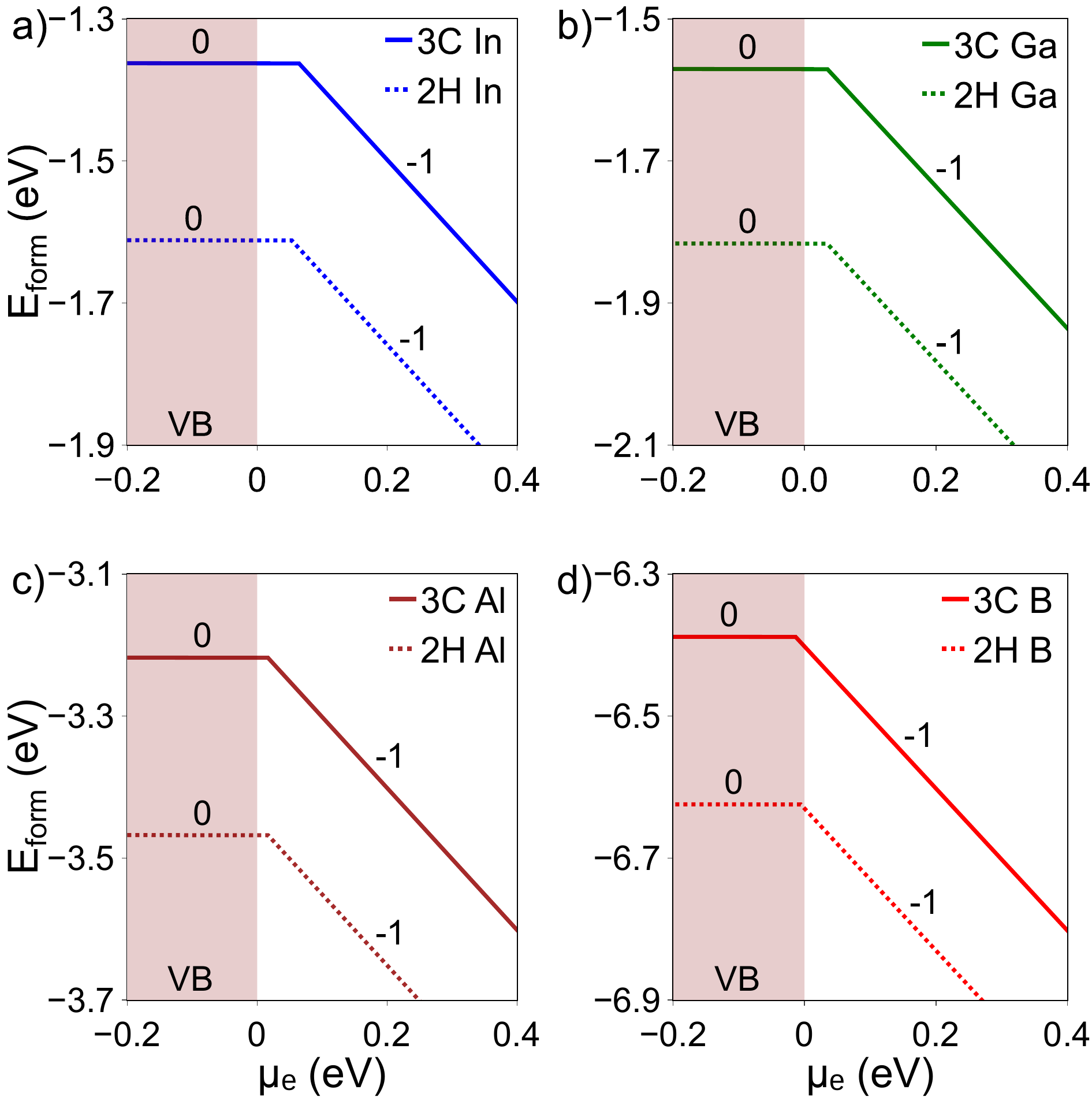} 
\caption{\label{fig:formation_energy_acceptors} Formation energy diagram as a function of the chemical potential for acceptors a) In, b) Ga, c) Al, and d) B. For each dopant the plots are reported for two values of $q$ (-1 and 0) and they are linear functions of the chemical potential as shown in Eq.~\ref{eq:formation_energy}. The chemical potential is set to 0 at the valence band maximum. The intersection of the formation energy lines for two charge states $q$~=~0 and $q$~=~-1 is the thermodynamic transition level calculated through Eq.~\ref{eq:transition_level}. The continuous line corresponds to the 3C phase and the dashed line to the 2H. The low energy dark pink region corresponds to the valence band.}
\end{figure}

\begin{figure}
\includegraphics[width=\columnwidth]{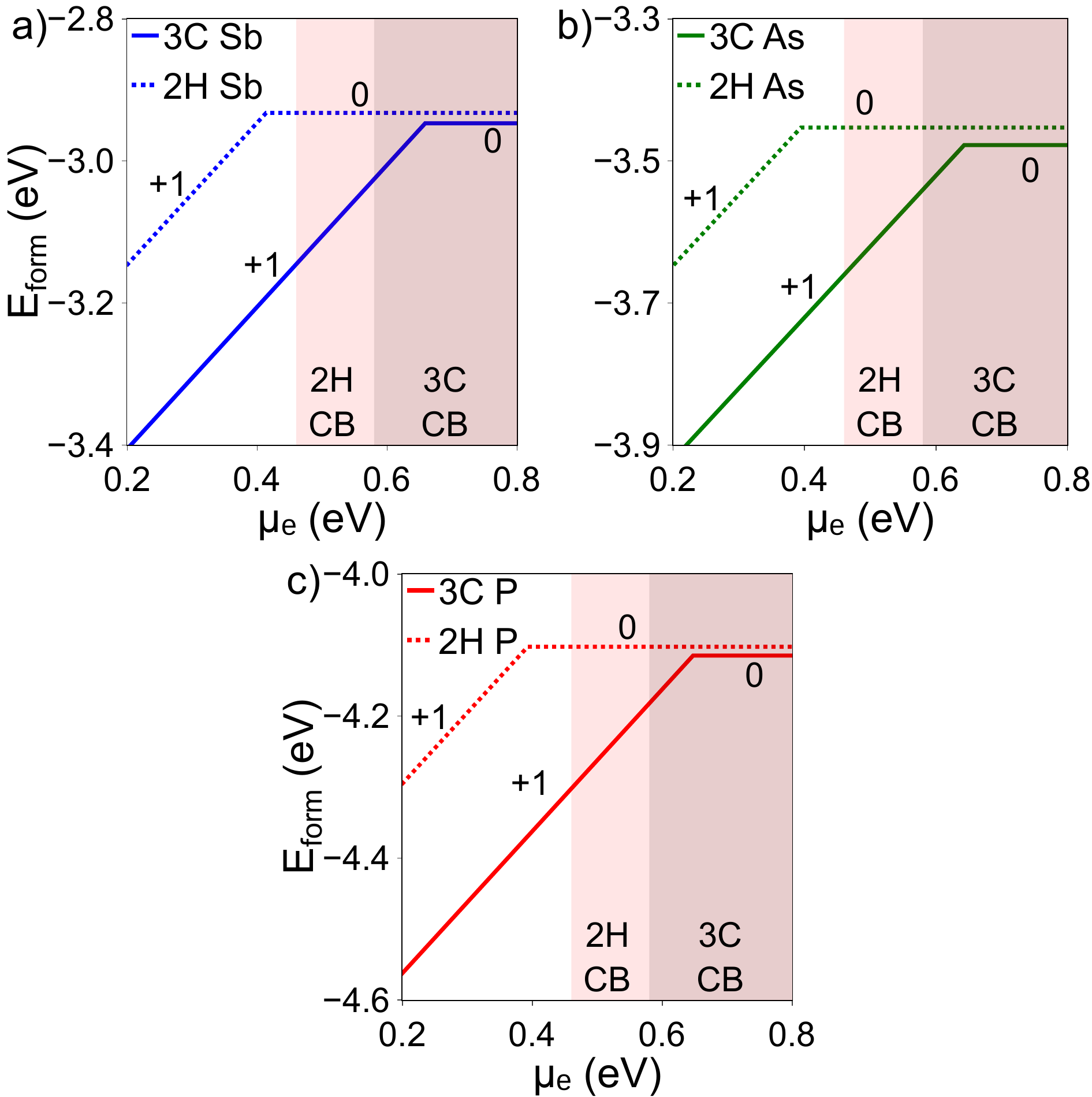} 
\caption{\label{fig:formation_energy_donors} Formation energy diagram as a function of the chemical potential for donors a) Sb, b) As, and c) P. For each dopant the plots are reported for two values of $q$ (0 and +1) and they are linear functions of the chemical potential as shown in Eq.~\ref{eq:formation_energy}. The chemical potential is set to 0 at the valence band maximum. The intersection of the formation energy lines for two charge states $q$~=~+1 and $q$~=~0 is the thermodynamic transition level calculated through Eq.~\ref{eq:transition_level}. The continuous line corresponds to the 3C phase and the dashed line to the 2H. The first high energy pink region corresponds to the hexagonal conduction band while the high energy pink region (darker) to the 3C conduction band.}
\end{figure}

The local structural analysis cannot provide alone a full picture of the impurity stability in different phases. Formation energy calculations of both neutral and charged dopants are required to analyze this question in depth. Figures~\ref{fig:formation_energy_acceptors} and \ref{fig:formation_energy_donors} report the formation energy, $E_{form}$, calculated with Eq.~\eqref{eq:formation_energy}, as a function of the chemical potential relative to the maximum of the valence band for p- and n-type dopants, respectively. Numerical values associated to the graphical plots of Figures~\ref{fig:formation_energy_acceptors} and \ref{fig:formation_energy_donors} are provided in Table~\ref{tab:acceptor_transition_levels_formation_energies} and Table~\ref{tab:donor_transition_levels_formation_energies}.

Looking at Figure~\ref{fig:formation_energy_acceptors}, one can see, as expected, that negatively charged acceptors are more stable than neutral impurities because of their tendency to attract an electron from the host crystal to create a hole in the valence band. Interestingly, for both the neutral and charged cases, their formation energy is always lower in the 2H phase with respect to the 3C one (see Figure~\ref{fig:formation_energy_acceptors}). This is even clearer if one looks at the first column of Table~\ref{tab:acceptor_transition_levels_formation_energies} where the difference in formation energy, $\Delta E^{2H-3C}$ for a given impurity in the 3C and in the 2H phase is reported for both the charged and the neutral cases. The more $\Delta E^{2H-3C}$ is negative, the more the impurity will prefer the 2H phase with respect to the 3C-one.  

This behavior can be ascribed to the clear preference for acceptors to be in the 2H phase as a consequence of its stronger tendency to induce a local $C_{3v}$ symmetry and a three-fold coordination around the impurity (see Section~\ref{sec:structural analysis}). The same explanation still holds when charged impurities are considered as this phase preference does not change (see Figure~\ref{fig:formation_energy_acceptors}).  

Similarly to the case of acceptors, Figure~\ref{fig:formation_energy_donors} shows that positively charged donors are more stable than neutral impurities because the additional donor electron, occupying an antibonding state, is only weakly bonded to the impurity and tends to be delocalized in the crystal. Furthermore, while neutral donors do not present any clear phase preference as $\Delta E^{2H-3C}$ is close to zero, charged donors reveal a marked tendency to be more stable in the 3C-phase (see Figure~\ref{fig:formation_energy_donors} and the first two columns of Table~\ref{tab:donor_transition_levels_formation_energies}). So for charged donors the 3C-host lattice is the ideal crystal environment to form four equal length covalent bonds with a marked $T_{d}$ symmetry.

Our results clearly show that, close to the 2H/3C homo-junction~\cite{VincentNL2015,TangNANOSCALE2017,AmatoNL2019}, the $\Delta E_{form}^{2H-3C}$ of charged donors is positive while for charged acceptors it is negative. To align the Fermi levels of the two phases a space charge will be generated and thus the dopants will be necessarly charged at the interface. It is hence expected that a gradient in the dopant concentration at the interface between the two phases can be build up with donors accumulating on the 3C-Si side and acceptors in 2H-Si.

\section{\label{sec:transition}Thermodynamic transition levels}
The calculation of the formation energies of neutral and charged impurities allows to employ Eq.~\eqref{eq:transition_level} to calculate thermodynamic transition levels, $\epsilon(q_{1}/q_{2})$, which are essential for the experimental characterization of the defect. These charge transition energies correspond to the crossing point where the charged and the neutral defect formation energies coincide, as observed in Figures~\ref{fig:formation_energy_acceptors} and~\ref{fig:formation_energy_donors}. It is clear from the figures that, for acceptor dopants, the difference in the defects formation energy between the two phases changes in the order of a few meV with the state of charge of the impurities ($\Delta \epsilon_{A}^{2H-3C}$ also reported in the sixth column of Table~\ref{tab:acceptor_transition_levels_formation_energies}). More negative values of $\Delta \epsilon_{A}^{2H-3C}$ correspond to a transition level that is shallower in the 2H phase. For B impurities the transition energy is positive, those of Ga and Al are practically zero (less than 2 meV), while that of In is negative. This means that by increasing the size of the defect, the difference between the transition energies between 2H and 3C will be more negative and hence the defect level will be shallower in 2H-Si. In the case of smaller radius impurities, $\Delta \epsilon_{A}^{2H-3C}$ is positive and the defect level is shallower in 3C-Si. 
In summary, if one looks at the second and sixth column of Table~\ref{tab:acceptor_transition_levels_formation_energies}, it is possible to conclude that charged acceptors are always more stable in the 2H phase and present charge transition levels shallower with respect to the 3C-Si. The magnitude of these energy shift depends from the ionic radius of the dopant.

A similar analysis can be performed for donors. The difference of the transition energies between the two phases is about hundreths of meV (see sixth column of Table~\ref{tab:donor_transition_levels_formation_energies} and Fig.~\ref{fig:formation_energy_donors}) and are always negative. This means that the donor dopant levels will be always shallower in 3C-Si with respect to 2H-Si. As in the case of acceptors, even here we can recognize a relation with the ionic radius of the dopant. Indeed, $\Delta \epsilon_{D}^{2H-3C}$ decreases as a function of the radius mismatch. These findings agree with the hydrogenic model of donors that considers four covalent bonds and an unpaired one whose energy relative to the bottom of the conduction band is directly proportional to the effective mass and inversely proportional to the dielectric function. As reported in Table~\ref{tab:table_pure_si}, the in-plane component of the dielectric function is lower for 2H- than for 3C-Si. This means that the transition level of donors in 2H-Si will be deeper with respect to those of 3C-Si. The discussed behavior, with the appropriate differences, was already observed in polytypes of GaAs, GaP, and InP~\cite{GiorgiJPC2020}.

Nevertheless a note of caution is needed when looking at the transition levels shown in Figure~\ref{fig:formation_energy_acceptors} and Figure~\ref{fig:formation_energy_donors}. Indeed, some of these levels are resonant with the crystal host band states. This is directly related to the use of total energies obtained from semi-local functionals which cannot provide accurate ionization energies for shallow dopants~\cite{WangJAP2009,ZhangPRL2013} but can only qualitatively reproduce the experimental chemical trend among different type of dopants, as clearly shown in the last column of Table~\ref{tab:acceptor_transition_levels_formation_energies} and Table~\ref{tab:donor_transition_levels_formation_energies}. More advanced methods, such as quasi-particle G$_0$W$_0$ calculations, are known to provide accurate results but they are associated with a very high computational cost. Although in our case the absolute values of the transition energies for each phase taken separately may not be correct, the calculated semi-quantitative variation between the two phases is still valid and insightful.
To further improve the comparison of the transition energy levels between the two phases, we calculated them with respect to a common reference point, i.e. the vacuum level, $\Phi$, calculated from a Si slab model~\cite{JunqueraPRB2003,Bertocchi2017} as described by Eq.~\eqref{eq:transition_level_phi} (see SM for more details about the procedure~\cite{supp}). Indeed, even when the theoretical description is improved with beyond-DFT methods, band edges in semiconductors undergo significant shifts but charge transition levels of defects remain practically unaffected with respect to their semi-local energy positions~\cite{AlkauskasPRB2011,KomsaPRB2010,AlkauskasPRL2008}. Thus the calculated $\Delta \epsilon_{A}^{2H-3C}$ and $\Delta \epsilon_{D}^{2H-3C}$ presented in Table~\ref{tab:acceptor_transition_levels_formation_energies} and Table~\ref{tab:donor_transition_levels_formation_energies} and graphically shown in Fig.~\ref{fig:transition_energies} are a valid reference for future experiments aiming to measure the variation in the ionization energy among the two phases. 
 
 In agreement with other studies~\cite{BelabbesPRB2022,KaewmarayaJPC2017,KellerPRM2023}, we found that the 2H/3C Si junction presents a type-II alignment, as shown by the difference between hexagonal and cubic valence band maximum (VBM), $\Delta VBM^{2H-3C} \approx 0.154$ eV and between hexagonal and cubic conduction band minimum (CBM), $\Delta CBM^{2H-3C} \approx 0.036$ eV. More accurate calculations~\cite{KellerPRM2023} with the Heyd-Scuseria-Ernzerhof hybrid functional (HSE06)~\cite{Heyd2003} and the modified Becke-Johnson potential combined with local-density approximation correlation (mBJLDA)~\cite{Tran2009} functional report a valence band offset of $\Delta VBM^{2H-3C}_{mBJLDA} = 0.26$~eV and $\Delta VBM^{2H-3C}_{HSE} = 0.31$~eV. Calculations for the hexagonal/cubic homojunction supercell using the Local Density Approximation (LDA) with the GW correction~\cite{AmatoNL2016} estimate a value of $\Delta VBM^{2H-3C}_{GW} = 0.25$~eV, slightly larger than the one we obtained. On the other hand, in the same work~\cite{AmatoNL2016} a value of $\Delta CBM^{2H-3C}_{GW} = 0.20$~eV was reported, while Keller et al.~\cite{KellerPRM2023} estimate conduction band offsets to be $\Delta CBM^{2H-3C}_{MBJLDA} = 0.07$~eV and $\Delta CBM^{2H-3C}_{HSE} = 0.04$~eV, which is closer to our GGA results. It is possible to conclude that, despite the use of different computational approaches and alignment procedures, the type-II character of the band offset is observed in various studies~\cite{AmatoNL2016,BelabbesPRB2022,KaewmarayaJPC2017,KellerPRM2023}. The 3C-VBM lies well below the 2H-VBM, with differences on the order of hundreds of meV, similar to the magnitude of the calculated $|\Delta E^{2H-3C}_{form,q}|$. When comparing the two phases, acceptor transition levels of both materials (referred to the vacuum level) are energetically lower in 3C-Si compared to those in 2H-Si. For donors, the band offset will lead to a closer alignment of the energy levels, although their exact relative positions may slightly differ slightly depending on the computational approach used. 

\section{\label{sec:bandoffset}Band-offset model of doping}
\begin{figure} 
\centering
\includegraphics[width=\columnwidth]{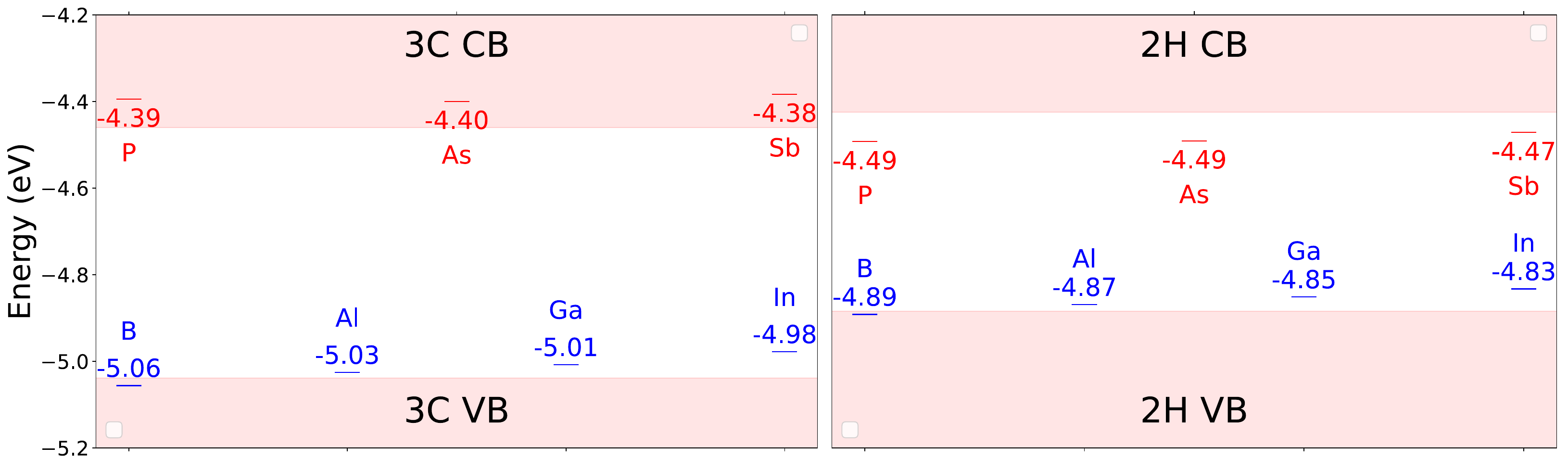}
\caption{\label{fig:transition_energies} Transition energy levels of acceptors, $\overline{\epsilon}_{A}$, (red lines) and donors, $\overline{\epsilon}_{D}$, (blue lines) in 3C-Si (left) and 2H-Si (right) as calculated in Eq.~\eqref{eq:transition_level_phi}. It is worth to note that $\overline{\epsilon}_{A}$ and $\overline{\epsilon}_{D}$ are evaluated referring to the vacuum potential $\Phi$ taken as zero of the energy. The DFT-GGA calculated value of ionization potential and electronic affinity for 3C-Si is in good agreement with previous theoretical works using the same approach~\cite{JiangJCP2013,MarriPCCP2020,Bertocchi2017}.}
\end{figure}

\begin{figure}
\includegraphics[width=0.7\columnwidth]{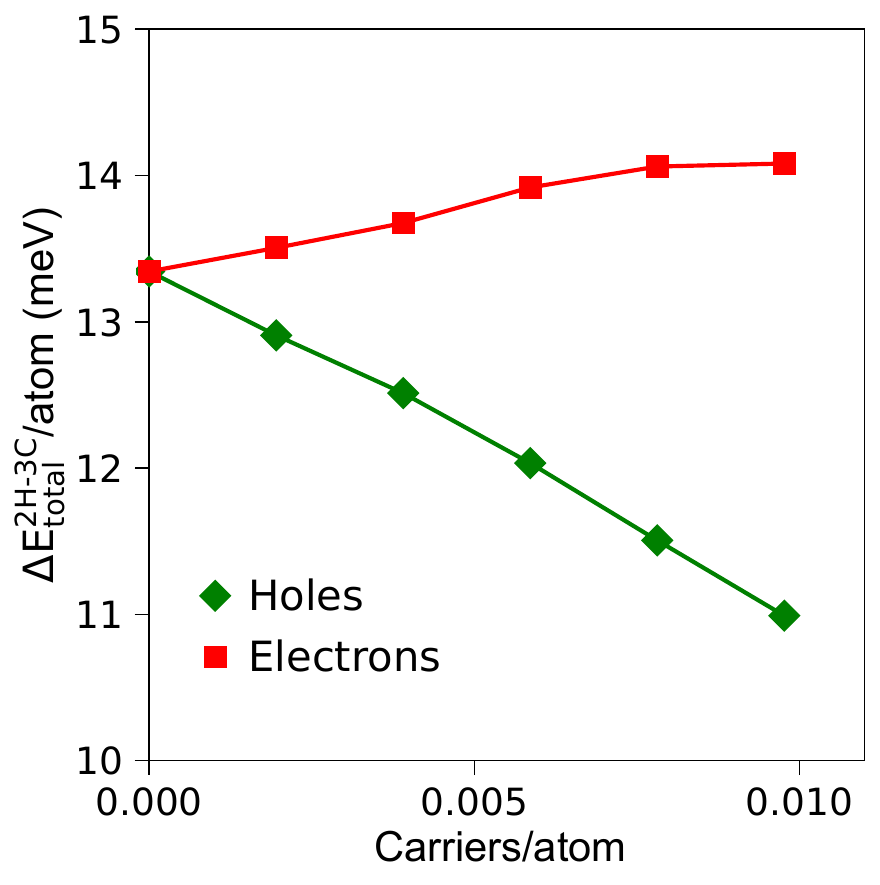}    
\caption{\label{fig:pure_doping} Difference in the total energy, $\Delta E_{total}^{2H-3C}$, per atom between 2H-Si and 3C-Si as a function of the carrier concentration per atom for electrons (red square symbols line) and hole (green diamond symbols line).}
\end{figure}

In order to further confirm our DFT findings and to depict a full physical picture of the dopant stability in 2H and 3C-Si, we build up a simple model following the arguments employed by Dalpian et al.~\cite{DalpianAPL2006,DalpianPRL2004} to explain the stability of III-V polytypes and recently adopted by Cai et al.~\cite{CaiPRB2021} to study 2H-Ge. Ideal doping can be represented by simply adding electrons and holes to the pristine supercell of 2H- and 3C-Si taking into account total energy correction due to finite size effects~\cite{FreysoldtPRL2009}. Figure~\ref{fig:pure_doping} reports the difference in the total energy between the two phases, $\Delta E_{total}^{2H-3C}$ per atom as a function of the carrier concentration. The results show that introducing electrons tends to stabilize the 3C phase while introducing holes is a way to make the 2H phase more stable. These results are a consequence of the type II band offset between the two phases: since 2H-Si has a higher valence band maximum than 3C (see Fig.~\ref{fig:transition_energies}), it will cost less energy to create holes in the 2H-phase than in the 3C-one. On the other hand, since 3C-Si has a lower conduction band minimum than 2H-Si (see Fig.~\ref{fig:transition_energies}), it will cost less energy to add electrons in the 3C-Si than in 2H-Si. Moreover, as the 2H/3C conduction band offset is very tiny, the slope of the electron doping curve (red line of Fig.~\ref{fig:pure_doping}) is found to be small while that of the hole incorporation is larger (green line of Fig.~\ref{fig:pure_doping}). Interestingly, this fully confirms our DFT results for impurity doped supercells which show that i) acceptors are more stable in the 2H phase and have shallower transition energy levels if compared to 3C-Si, ii) donors prefer instead to occupy cubic crystal site with shallower transition energies if compared to 2H-Si.  

\section{Conclusion}
In conclusion, this work presents DFT calculations of the structural and electronic properties of p- and n-type impurities in 2H-Si, comparing them with the 3C-Si case. We find that acceptors tend to be more stable in the hexagonal phase because of its stronger tendency to induce a local $C_{3v}$ symmetry and three-fold coordination around the impurity. This behaviour is valid for all the considered acceptors (B, Al, Ga and In). In particular, charged acceptors are always more stable in the 2H phase and present a shallower charge transition level with respect to the 3C-Si crystal host.

On the other hand, we find that neutral donors (P, As, Sb) have no phase preference while positively charged donors have lower formation energies in the cubic phase than in the hexagonal one. This can be ascribed to the fact that for charged donors the 3C-host lattice is the ideal crystal environment to form four equal length covalent bonds with a marked $T_{d}$ symmetry. Furthermore, donors levels are shallowers in 3C-Si than in 2H-Si. 

The physical validity of the doped supercell calculations is reinforced by a simple model based on the 2H/3C band offset diagram through which we show how holes can be used to stabilize the 2H-Si phase while electrons prefer the 3C-Si crystal phase.

All these results assume a particular importance for more accurate calculations beyond the DFT framework and for future experiments. On the one hand, these results could be further assessed by hybrid-DFT or GW based calculations that could confirm the difference in transition level positions among 2H and 3C-Si by providing numerical values to be directly compared with experiments. On the other hand, being a theoretical estimation of transition energy levels in 2H-Si, these findings represent a valid theoretical reference for future measurements of the variation in the ionization energy of a given dopant among the two phases. Moreover, in the case of 2H/3C homo-junctions~\cite{VincentNL2015,TangNANOSCALE2017,AmatoNL2019}, because of the different stability of charged dopants, a gradient in dopant concentration at the interface between the two phases may be experimentally detected with donors accumulating on the 3C-Si side and acceptors in 2H-Si.

\begin{acknowledgments}
All the authors acknowledge the ANR AMPHORE project (ANR-21-CE09-0007) and the ANR TULIP (ANR-24-CE09-5076). Part of the high-performance computing resources for this project were granted by the Institut du d\'eveloppement et des ressources en informatique scientiﬁque (IDRIS) under the allocations AD010914974 and AD010915077 via GENCI (Grand Equipement National de Calcul Intensif). The authors also acknowledge the use of the CERES high performance computer cluster at the Laboratoire de Physique des Solides of the Université Paris-Saclay (Orsay, France).
\end{acknowledgments}

\bibliography{manuscript}

\pagebreak
\newpage
\begin{table}
\caption{\label{tab:table_pure_si} Calculated GGA lattice parameter, bandgap energy, and static dielectric constants components for 3C- and 2H-Si as well as the corresponding experimental values.}
\begin{ruledtabular}
\begin{tabular}{ccccccc}
&a ({\AA}) & c ({\AA}) &$E_{GAP}$ (eV)& $\epsilon_{xx}$&$\epsilon_{zz}$ \\ \hline
3C-Si & 5.45 (5.43\footnotemark[1]) & / & 0.58 (1.12-1.23\footnotemark[2]) & 13.07 (11.7\footnotemark[3]) & 13.07 (11.7\footnotemark[3]) \\
2H-Si & 3.84 (3.82-3.84\footnotemark[4]) & 6.34 (6.26-6.34\footnotemark[4]) & 0.46 (0.92\footnotemark[5]) & 12.89 & 13.90 \\
\end{tabular}
\end{ruledtabular}
\footnotetext[1]{From Ref.~\cite{Isherwood1966} and \cite{Godwod1974}.}
\footnotetext[2]{From Ref.~\cite{Precker2002} and \cite{Low2008}.}
\footnotetext[3]{From Ref.~\cite{Li1980}.}
\footnotetext[4]{From Ref.~\cite{HaugeNL2015}, \cite{Ahn2021}, \cite{Zhang1999}, and~\cite{Besson1987}. Values summarized in Ref.~\cite{KellerPRM2023}.}
\footnotetext[5]{From Ref.~\cite{Fabbri2014}.}
\end{table}

\pagebreak
\newpage

\begin{table}
\sisetup{
  round-mode=places,
  round-precision=3,
  table-format=1.2
}
\caption{\label{tab:acceptors_geometry} Impurity first neighbor's bond distances for acceptors in the 3C-Si and 2H-Si host lattices after relaxation. For the 2H phase $d_1$, $d_2$, and $d_3$ correspond to distances calculated in the c-(0001) plane while $d_4$ lies along the [0001] hexagonal direction. In parenthesis the contraction (negative) or expansion (positive) with respect to the pristine Si distances are given.}
\begin{ruledtabular}
\begin{adjustbox}{width=0.8\textwidth}
\begin{tabular}{cccccc}
 &  &$d_{1}$ (\AA)&$d_{2}$ (\AA)&$d_{3}$ (\AA)&$d_{4}$ (\AA)  \\ \hline
\multirow{4}{*}{B} & \multirow{2}{*}{3C}
&\num{2.079}&\num{2.079}&\num{2.079}&\num{2.079}   \\
 & & (-11.9\%)
 & (-11.9\%)
 & (-11.9\%)
 & (-11.9\%)\\
& \multirow{2}{*}{2H}&\num{2.076}&\num{2.076}&\num{2.076}&\num{2.092} \\  
& & (-11.9\%)
& (-11.9\%)
& (-11.9\%)
& (-11.8\%) \\  \hline
\multirow{4}{*}{B$^{-}$} & \multirow{2}{*}{3C}&\num{2.079}&\num{2.079}&\num{2.079}&\num{2.079}   \\
& & (-11.9\%) 
& (-11.9\%) 
& (-11.9\%) 
& (-11.9\%) \\
& \multirow{2}{*}{2H}&\num{2.073}&\num{2.073}&\num{2.073}&\num{2.095}\\ 
& & (-12\%)
& (-12\%)
& (-12\%)
& (-11.6\%) \\
\hline \hline
\multirow{4}{*}{Al} & \multirow{2}{*}{3C}
&\num{2.427}&\num{2.427}&\num{2.427}&\num{2.427}   \\
& &	(2.9\%) 
&	(2.9\%) 
&	(2.9\%) 
&	(2.9\%)
\\
& \multirow{2}{*}{2H}&\num{2.420}&\num{2.427}&\num{2.427}&\num{2.430} \\ 
& &	(2.8\%)
&	(3.0\%)
&	(3.0\%)
&	(2.5\%) \\ \hline
\multirow{4}{*}{Al$^{-}$} & \multirow{2}{*}{3C}&\num{2.426}&\num{2.426}&\num{2.426}&\num{2.426} \\
& &	(2.8\%)
&	(2.8\%)
&	(2.8\%)
&	(2.8\%)
 \\
& \multirow{2}{*}{2H}&\num{2.421}&\num{2.421}&\num{2.421}&\num{2.433} \\ 
& &	(2.8\%)
&	(2.8\%)
&	(2.8\%)
&	(2.6\%)
 \\
\hline \hline
\multirow{4}{*}{Ga} & \multirow{2}{*}{3C}
&\num{2.470}&\num{2.470}&\num{2.470}&\num{2.470}   \\
& &	(4.7\%)
&	(4.7\%)
&	(4.7\%)
&	(4.7\%)
 \\
& \multirow{2}{*}{2H}&\num{2.461}&\num{2.473}&\num{2.473}&\num{2.463} \\ 
& &	(4.5\%)
&	(5.0\%)
&	(5.0\%)
&	(3.9\%)
 \\ \hline
\multirow{4}{*}{Ga$^{-}$} & \multirow{2}{*}{3C}&\num{2.464}&\num{2.464}&\num{2.464}&\num{2.464}\\
& &	(4.4\%)
&	(4.4\%)
&	(4.4\%)
&	(4.4\%)
 \\
& \multirow{2}{*}{2H}&\num{2.462}&\num{2.463}&\num{2.463}&\num{2.466} \\ 
& &	(4.6\%)
&	(4.6\%)
&	(4.6\%)
&	(4.0\%)
 \\
\hline \hline
\multirow{4}{*}{In} & 3C&\num{2.581}&\num{2.581}&\num{2.581}&\num{2.581} \\
& &	(9.4\%)
&	(9.4\%)
&	(9.4\%)
&	(9.4\%)
 \\
& \multirow{2}{*}{2H}&\num{2.572}&\num{2.586}&\num{2.586}&\num{2.569} \\ 
& &	(9.2\%)
&	(9.8\%)
&	(9.8\%)
&	(8.4\%)
\\ \hline
\multirow{4}{*}{In$^{-}$} & \multirow{2}{*}{3C}&\num{2.571}&\num{2.571}&\num{2.571}&\num{2.571}  \\
& &	(9.0\%)
&	(9.0\%)
&	(9.0\%)
&	(9.0\%)

 \\
& \multirow{2}{*}{2H}&\num{2.571}&\num{2.573}&\num{2.573}&\num{2.573} \\ 
& &	(9.2\%)
&	(9.2\%)
&	(9.2\%)
&	(8.5\%)
 \\
\end{tabular}
\end{adjustbox}
\end{ruledtabular}
\end{table}

\pagebreak
\newpage

\begin{table}
\sisetup{
  round-mode=places,
  round-precision=2,
  table-format=1.2
}
\caption{\label{tab:donors_geometry} Impurity first neighbor's bond distances for donors in the 3C-Si and 2H-Si host lattices after relaxation. For the 2H phase $d_1$, $d_2$, and $d_3$ correspond to distances calculated in the c-(0001) plane while $d_4$ lies along the [0001] hexagonal direction. In parenthesis the contraction (negative) or expansion (positive) with respect to the pristine Si distances are given.}
\begin{ruledtabular}
\begin{tabular}{cccccc}
 &  &$d_{1}$ (\AA)&$d_{2}$ (\AA)&$d_{3}$ (\AA)&$d_{4}$ (\AA)  \\ \hline
\multirow{4}{*}{P} & \multirow{2}{*}{3C}& \num{2.358}& \num{2.358}& \num{2.358}& \num{2.358}  \\
&&	(-0.1~\%)
&	(-0.1~\%)
&	(-0.1~\%)
&	(-0.1~\%)\\ 

& \multirow{2}{*}{2H}&\num{2.348}&\num{2.350}&\num{2.350}&\num{2.470} \\  
&&	(-0.2~\%)
&	(-0.2~\%)
&	(-0.2~\%)
&	(-0.1~\%)\\
\hline
\multirow{4}{*}{P$^{+}$} & \multirow{2}{*}{3C}&\num{2.356}&\num{2.356}&\num{2.356}&\num{2.356}   \\
&&	(-0.1~\%)
&	(-0.1~\%)
&	(-0.1~\%)
&	(-0.1~\%) \\
& \multirow{2}{*}{2H}&\num{2.349}&\num{2.349}&\num{2.349}&\num{2.370} \\ 
&&	(-0.2~\%)
&	(-0.2~\%)
&	(-0.2~\%)
&	(0.0~\%)\\
\hline
\hline
\multirow{4}{*}{As} &  \multirow{2}{*}{3C}&\num{2.470}&\num{2.470}&\num{2.470}&\num{2.470}  \\
&&	(4.7~\%)
&	(4.7~\%)
&	(4.7~\%)
&	(4.7~\%)\\

&  \multirow{2}{*}{2H}&\num{2.435}&\num{2.436}&\num{2.436}&\num{2.452} \\ 
&&	(5.0~\%)
&	(5.0~\%)
&	(5.0~\%)
&	(3.9~\%)\\
\hline
\multirow{4}{*}{As$^{+}$} &  \multirow{2}{*}{3C}&\num{2.464}&\num{2.464}&\num{2.464}&\num{2.464}  \\
&&	(4.4~\%)
&	(4.4~\%)
&	(4.4~\%)
&	(4.4~\%)\\
&  \multirow{2}{*}{2H}&\num{2.433}&\num{2.433}&\num{2.433}&\num{2.451} \\ 
&&	(4.6~\%)
&	(4.6~\%)
&	(4.6~\%)
&	(4.0~\%)\\
\hline
\hline
\multirow{4}{*}{Sb} & \multirow{2}{*}{3C}
&\num{2.574}&\num{2.574}&\num{2.574}&\num{2.574}  \\

&&	(9.1~\%)
&	(9.1~\%)
&	(9.1~\%)
&	(9.1~\%) \\
& \multirow{2}{*}{2H}&\num{2.570}&\num{2.570}&\num{2.570}&\num{2.582} \\ 
&&	(9.1~\%)
&	(9.1~\%)
&	(9.1~\%)
&	(8.9~\%) \\ \hline
\multirow{4}{*}{Sb$^{+}$} & \multirow{2}{*}{3C}&\num{2.573}&\num{2.573}&\num{2.573}&\num{2.573}   \\
&&	(9.0~\%)
&	(9.0~\%)
&	(9.0~\%)
&	(9.0~\%)\\
& \multirow{2}{*}{2H}&\num{2.570}&\num{2.570}&\num{2.570}&\num{2.582} \\
&&	(9.1~\%)
&	(9.1~\%)
&	(9.1~\%)
&	(8.9~\%)
\end{tabular}
\end{ruledtabular}
\end{table}

\pagebreak
\newpage 

\pagebreak
\newpage
 \begin{table*}[htbp]
\centering
\caption{\label{tab:acceptor_transition_levels_formation_energies} Acceptors: columns two and three report the difference in the formation energy between the 2H and 3C phases for neutral ($\Delta E^{\text{2H-3C}}_{0, \text{form}}$) and charged impurities ($\Delta E^{\text{2H-3C}}_{-1, \text{form}}$). The third and fourth columns correspond to the transition energy levels for the 2H and 3C phases calculated Eq.~\eqref{eq:transition_level} between the charge state $q$~=~0 and $q$~=~-1. The fifth column reports the difference in acceptor transition energies, $\Delta \epsilon^{\text{2H-3C}}_A$, between the 2H and 3C phases. Negative (positive) values correspond to shallower (deeper) states in the hexagonal phase. The last column lists the experimental transition levels for the cubic phase.}
\begin{tabular}{cccccccc}
\hline \hline
& $\Delta E^{\text{2H-3C}}_{0, \text{form}}$ & $\Delta E^{\text{2H-3C}}_{-1, \text{form}}$ & $\epsilon^{\text{2H}}_A (0/-1)$ & $\epsilon^{\text{3C}}_A (0/-1)$ & $\Delta \epsilon^{\text{2H-3C}}_A$ & $\epsilon^{\text{3C, exp}}_A (0/-1)$ \\

& (eV) & (eV) & (meV) & (meV) & (meV) & (meV)\\
\hline
B & -0.24 & -0.23 & -7& -15& 8 & 30, 42-45 \footnotemark[1] &\\
Al & -0.25 & -0.25 & 16& 15 & 0 & 55-57 \footnotemark[2] & \\
Ga & -0.25 & -0.25 & 33& 34 & 0 & 58-65\footnotemark[3] & \\
In & -0.25 & -0.26 & 52& 63 & -11 & 150–180 \footnotemark[4] &  \\ \hline \hline
\end{tabular}
\footnotetext[1]{From Ref.~\cite{BursteinJPCS1956, Steger2009, Karaiskaj2003, Xiao2017}.}
\footnotetext[2]{From Ref.~\cite{BursteinJPCS1956, Vinh2013}.}
\footnotetext[3]{From Ref.~\cite{BursteinJPCS1956, Vinh2013}.}
\footnotetext[4]{From Ref.~\cite{BursteinJPCS1956, Fargi2014, Parker1983,Cerofolini1985,Jones1981}.}
\end{table*}

\pagebreak
\newpage

\pagebreak
\newpage

 \begin{table*}[htbp]
\centering
\caption{\label{tab:donor_transition_levels_formation_energies} Donors: columns two and three report the difference in the formation energy between the 2H and 3C phases for neutral ($\Delta E^{\text{2H-3C}}_{0, \text{form}}$) and charged impurities ($\Delta E^{\text{2H-3C}}_{+1, \text{form}}$). The third and fourth columns correspond to the transition energy levels for the 2H and 3C phases calculated with Eq.~\eqref{eq:transition_level} between the charge state $q$~=~+1 and $q$~=0. The fifth column reports the difference in donors transition energies, $\Delta \epsilon^{\text{2H-3C}}_D$, between the 2H and 3C phases. Positive (negative) values correspond to shallower (deeper) states in the hexagonal phase. The last column lists the experimental transition levels for the cubic phase.}
\begin{tabular}{cccccccc}
\hline
& $\Delta E^{\text{2H-3C}}_{0, \text{form}}$ & $\Delta E^{\text{2H-3C}}_{+1, \text{form}}$ & $\epsilon^{\text{2H}}_D (+1/0)$ & $\epsilon^{\text{3C}}_D (+1/0)$& $\Delta \epsilon^{\text{2H-3C}}_D$ & $\epsilon^{\text{3C, exp}}_D (+1/0)$\\
& (eV) &(eV) &(meV) &(meV) &(meV) & (meV)\\\hline
P& 0.01 & 0.27 & 392 & 647 & -145 & 1076 – 1078  \footnotemark[1]  \\
As & 0.02 & 0.27 & 394 & 642 & -138 & 1071  \footnotemark[2]   \\
Sb & 0.01 & 0.26 & 413 & 658 & -135 & 1081  \footnotemark[3]   \\ \hline \hline
\end{tabular}
\footnotetext[1]{From Ref.~\cite{Picus1956,Vinh2013,Karaiskaj2003,Xiao2017,Steger2009}.}
\footnotetext[2]{From Ref.~\cite{Picus1956}.}
\footnotetext[3]{From Ref.~\cite{Picus1956,Vinh2013}.}
\end{table*}

\pagebreak
\newpage

\pagebreak
\newpage

\pagebreak

\end{document}